\documentclass[english,manuscript=article,layout=twocolumn]{achemso}

\usepackage[T1]{fontenc}
\usepackage[latin9]{inputenc}
\usepackage{geometry}
\usepackage{xcolor}
\usepackage{pdfcolmk}
\usepackage{babel}
\usepackage{array}
\usepackage{booktabs}
\usepackage{textcomp}
\usepackage{amsmath}
\usepackage{graphicx}
\PassOptionsToPackage{normalem}{ulem}
\usepackage{ulem}
\usepackage[unicode=true,
    bookmarks=true,
    bookmarksnumbered=false,
    bookmarksopen=false, 
    breaklinks=true, 
    pdfborder={0 0 0},
    pdfborderstyle={}, 
    backref=false, 
    colorlinks=true,
    allcolors=blue]{hyperref}
\hypersetup{pdftitle={},
 pdfauthor={Sohvi Luukkonen}}

\title{Predicting hydration free energies of the FreeSolv database  of druglike molecules with molecular density functional theory}
\author{Sohvi Luukkonen}
\affiliation{Maison de la Simulation, CNRS-CEA-Universit{\'e} Paris-Saclay,
91191 Gif-sur-Yvette, France}
\author{Luc Belloni}
\affiliation{LIONS, NIMBE, CEA, CNRS, Universit{\'e} Paris-Saclay, 91191 Gif-sur-Yvette, France}
\author{Daniel Borgis}
\affiliation{Maison de la Simulation, CNRS-CEA-Universit{\'e} Paris-Saclay,
91191 Gif-sur-Yvette, France}
\alsoaffiliation{PASTEUR, D{\'e}partement de chimie, {\'E}cole normale sup{\'e}rieure, PSL University,
Sorbonne Universit{\'e}, CNRS, 75005 Paris, France}
\author{Maximilien Levesque}
\affiliation{PASTEUR, D{\'e}partement de chimie, {\'E}cole normale sup{\'e}rieure, PSL University,
Sorbonne Universit{\'e}, CNRS, 75005 Paris, France}
\alsoaffiliation{Aqemia, Paris, France}
\email{*maximilien.levesque@aqemia.com}
\providecommand{\tabularnewline}{\\}
\providecolor{lyxadded}{rgb}{0,0,1}
\providecolor{lyxdeleted}{rgb}{1,0,0}

\setkeys{acs}{articletitle=true,doi=true}
\newcommand*{\doi}[1]{\href{http://dx.doi.org/#1}{#1}}

\begin{document}
\begin{abstract}
We assess the performance of molecular density functional theory (MDFT)
to predict hydration free energies of the small drug-like molecules benchmark, FreeSolv. MDFT
in the hyper-netted chain approximation (HNC) coupled with a pressure
correction predicts experimental hydration free energies of the FreeSolv
database within 1 kcal/mol with an average computation time of two
cpu.min per molecule. This is the same accuracy as for simulation
based free energy calculations that typically require hundreds of
cpu.h or tens of gpu.h per molecule.
\end{abstract}

\section{Introduction}

The ability to predict accurately solvation free energies (SFEs) and
solvation profiles unlocks the access to several key thermodynamical
quantities of biomolecular systems. SFEs, eventually combined with
gas-phase calculations, enable the computation of relative solubilities,
binding \citep{snyder_mechanism_2011,wang_ligand_2011} and transfer\citep{bannan_calculating_2016}
free energies, partition\citep{bannan_calculating_2016} or activity
coefficients. The computation of SFEs, i.e. the reversible work to
bring a molecule from vacuum to solvent is non-trivial as it requires
the sampling of all possible states that can be visited during the
transformation
In 2015, considering the difficulty\citep{skyner_2015}
but nevertheless necessity of evaluating precisely SFEs in the drug
design process, important actors of the pharmaceutical industry publicly
called the academic world for alternatives, pointing out the lack
of precision of current methods\citep{sherborne_collaborating_2016}.

Multiple approaches for SFE calculations\citep{skyner_2015} have been
developed since the early 20th century from multiple continuum models\citep{tomasi_molecular_1994,cramer_1999,tomasi_quantum_2005},
like COSMO-RS\citep{klamt_1993,klamt_1995,klamt_cosmo-rs_2016} or
AquaSol\citep{koehl_2010}, exact but resource consuming free energy
perturbation approaches with molecular simulations 
(MD+FEP) to well developed end-point approaches like WaterMap\citep{young_motifs_2007,Abel_2008}
or GIST\citep{nguyen_grid_2012}.

An alternative to these methods lies within liquid state theories\citep{hansen_theory_2013}.
Using liquid state solvation theories like molecular density functional
theory\citep{gendre_classical_2009,zhao_molecular_2011,borgis_molecular_2012,ding_2017} (MDFT) or 3D-RISM\citep{hirata_extended_1981,beglov_integral_1997,kovalenko_three-dimensional_1998,kovalenko_potential_1999}
, one can predict hydration free energies (HFE) and solvation structures
of complex solutes like proteins \citep{imai_solvation_2004,ding_2017,omelyan_2015,nguyen_2019},
aluminosilicate surfaces\citep{levesque_solvation_2012} or inorganic
complexes\citep{Ruankaew_2019}. At their heart lies the molecular
Ornstein-Zernike equation (MOZ)\citep{blum_invariant_1972,blum_invariant_1972-1}.
In the RISM approach, MOZ loses its molecular nature since molecular correlations are approximated by site-site correlations. The MDFT, on the other hand,
keeps the full molecular description of the solvent by solving the
full angular-dependent MOZ equation. Since Ding et al. \citep{ding_2017},
MDFT can be solved efficiently at the hyper-netted chain (HNC) approximation
level and predicts SFEs and equilibrium solvent structures in a few
minutes which is 3-5 orders of magnitude faster than MD+FEP calculations.

This paper aims at assessing the performance of MDFT-HNC coupled with
the recent van der Walls pressure correction\citep{robert_2020} to
predict HFEs of small drug-like molecules of the FreeSolv database\citep{matos_2017_freesolv,mobley_small_2009}.
The FreeSolv database contains HFEs obtained by experiments and state-of-the-art
MD+FEP calculations for 642 small neutral organic molecules. The database
has more than 70 chemical functions and molecular masses ranging from
16 to 493 Da. These are typical sizes for drug-like molecules as defined
by Lipinski's 'rule of five' \citep{lipinski_1997} of molecule's
drug-likeness with a molecular mass criteria maximum at 500 Da. Moreover,
95\% of the database has a molecular mass less than 300 Da defined
as a limit for lead-like molecules by the 'rule of three'\citep{congreve_2003}.
The database contains only neutral molecules as measuring SFEs of
an isolated charged species requires extra thermodynamic assumptions
or introduces other complexities\citep{matos_2017_freesolv} that are still not well understood.

In the first section, we present
briefly the MDFT framework and computational details. We refer the
reader to refs. \citep{gendre_classical_2009,zhao_molecular_2011,borgis_molecular_2012,ramirez_density_2002,jeanmairet_molecular_2013,jeanmairet_molecular_2013-1,ding_2017}
for a complete review. In the second section, we evaluate the capacity
of MDFT to predict experimental HFEs on the FreeSolv database, compare
these results to MD+FEP and 3D-RISM and do an error analysis
on selected features of the drug-like molecules in order to be able
to infer error bars on the method. Conclusions and perspectives are
presented in the third section. 


\section{Theory}

\begin{table*}[h!]
\begin{tabular}{c>{\centering}p{2.2cm}>{\centering}p{2.2cm}>{\centering}p{2.2cm}>{\centering}p{2.2cm}>{\centering}p{2.2cm}>{\centering}p{2.2cm}}
 & \multicolumn{3}{c}{{\small{}Full set (619)}} & \multicolumn{3}{c}{{\small{}Rigid sub-set (520})}\tabularnewline
\cmidrule{2-7} 
 & {\small{}MD+FEP$^{(a)}$} & {\small{}3D-RISM$^{(b)}$} & {\small{} MDFT$\mathrm{^{HNC+vdW}}$} & {\small{}MD+FEP} & {\small{}3D-RISM} & {\small{} MDFT$\mathrm{^{HNC+vdW}}$}\tabularnewline
\midrule
{\small{}MAE (kcal/mol)} & {\small{}$1.06\pm0.08$} & {\small{}$1.11\pm0.08$} & {\small{}$1.07\pm0.08$} & {\small{}$0.98\pm0.07$} & {\small{}$1.04\pm0.09$} & {\small{}$0.92\pm0.07$}\tabularnewline
{\small{}RMSE (kcal/mol)} & {\small{}$1.41\pm0.12$} & {\small{}$1.52\pm+0.13$} & {\small{}$1.49\pm0.13$} & {\small{}$1.29\pm0.11$} & {\small{}$1.45\pm0.14$} & {\small{}$1.25\pm0.11$}\tabularnewline
{\small{}ME (kcal/mol)} & {\small{}$-0.37\pm0.11$} & {\small{}$-0.15\pm0.12$} & {\small{}$0.07\pm0.12$} & {\small{}$-0.40\pm0.11$} & {\small{}$-0.19\pm0.13$} & {\small{}$-0.07\pm0.11$}\tabularnewline
{\small{}Max err. (kcal/mol)} & {\small{}$7.39$} & {\small{}$7.11$} & {\small{}$8.81$} & {\small{}$4.57$} & {\small{}$7.11$} & {\small{}$4.82$}\tabularnewline
{\small{}Pearson's $R$} & {\small{}$0.94\pm0.01$} & {\small{}$0.92\pm0.02$} & {\small{}$0.93\pm0.01$} & {\small{}$0.94\pm0.02$} & {\small{}$0.91\pm0.02$} & {\small{}$0.93\pm0.01$}\tabularnewline
{\small{}Spearman's $\rho$} & {\small{}$0.94\pm0.01$} & {\small{}$0.90\pm0.02$} & {\small{}$0.93\pm0.02$} & {\small{}$0.94\pm0.01$} & {\small{}$0.89\pm0.03$} & {\small{}$0.93\pm0.02$}\tabularnewline
{\small{}Kendall's $\tau$} & {\small{}$0.80\pm0.02$} & {\small{}$0.75\pm0.03$} & {\small{}$0.78\pm0.02$} & {\small{}$0.79\pm0.02$} & {\small{}$0.73\pm0.03$} & {\small{}$0.78\pm0.03$}\tabularnewline
{\small{}cpu.h per solute} & {\small{}$\sim10^{2}$} & {\small{}$\sim10^{-1}$} & {\small{}$\sim10^{-2}$} & {\small{}$\sim10^{2}$} & {\small{}$\sim10^{-1}$} & {\small{}$\sim10^{-2}$}\tabularnewline
\bottomrule
\end{tabular}

\caption{Summary of the statistical measures characterizing the correlations
between experimental HFEs and those obtained with simulation based
free energy techniques, 3D-RISM-KH and MDFT-HNC calculations for the
full FreeSolv database and a sub-set of rigid molecules. ME, MAE and
RMSE stand for mean, absolute and root-mean-squared error, respectively
and Pearson's $R$ is a linear correlation coefficient and Spearman's
$\rho$ and Kendall's $\tau$ are monotonic correlation coefficients.
All error bars correspond to the 95\% confidence interval\citep{errorbars}.
(a) Duarte Ramos Matos et al.\citep{matos_2017_freesolv} (b) Roy
and Kovalenko \citep{roy_2019}.\label{tab:Summary-of-the}}
\end{table*}

The molecular density functional theory of classical molecular fluids
computes the solvation free energy and equilibrium solvent density
around a solute. The solvation free energy of a given solute can be
defined as the difference between the grand potential $\Omega$ of
the solvated system and the grand potential $\Omega_{\mathrm{bulk}}$
of the bulk solvent. In the classical density functional framework\citep{mermin_thermal_1965,evans_nature_1979}
this difference can be expressed in a functional form: 
\begin{alignat}{1}
\Delta G_{\textrm{{solv}}}=\Omega-\Omega_{\mathrm{bulk}}=\underset{\rho\rightarrow\rho_{eq}}{\min}\left\{ \mathcal{F}[\rho]\right\} =\mathcal{F}[\rho_{\text{eq}}],\label{eq:minimisation}
\end{alignat}
where $\mathcal{F}[\rho]$ is a free energy functional to be minimized,
$\rho\equiv\rho(\boldsymbol{r},\omega)$ the molecular solvent density,
with $\boldsymbol{r}$ a three dimensional vector and $\omega$ the
Euler angles $(\theta,\phi,\psi)$, characterizing the position and
the orientation of the rigid solvent molecule relative to the rigid
solute, and $\rho_{\mathrm{eq}}$ the equilibrium solvent density.
In the absence of solute, the equilibrium density is the
homogeneous angular and spatial bulk density $\rho_{\textrm{bulk}}\equiv n_{\textrm{bulk}}i/8\pi\text{\texttwosuperior}$
where $n_{\textrm{bulk}}$ is the spatial homogeneous bulk density, typically 0.033~molecule per \AA$^{3}$ for water at room conditions ($\equiv1$~kg/L), and $i/8\pi^{2}$ is the angular normalization constant with $i$ the order of the main symmetry axis of the solvent molecule ($i=2$ for water which has
a $C_{\mathrm{2v}}$ symmetry with all the integrals of $\psi$ calculated implicitly between 0 and $\pi$).

Without approximations, we split the functional $\mathcal{F}$ into three parts: 
\begin{equation}
\mathcal{F}=\mathcal{F}_{{\rm id}}+\mathcal{F}_{{\rm ext}}+\mathcal{F}_{{\rm exc}},\label{eq:Fid+Fext+Fexc}
\end{equation}
where $\mathcal{F}_{{\rm id}}$ is the ideal term of a fluid of non-interacting
particles, $\mathcal{F}_{\textrm{ext}}$ is the external term induced
by the solute, and $\mathcal{F}_{{\rm exc}}$ is the excess term that
includes structural correlations between solvent molecules. 

The ideal term reads
\begin{equation}
\mathcal{F}_{{\rm id}}=k_{{\rm B}}T\int\mathrm{d}\mathbf{r}\mathrm{d}\mathbf{\omega}\left[\rho(\mathbf{r},\mathbf{\omega})\ln\left(\dfrac{\rho(\mathbf{r},\mathbf{\omega})}{\rho_{{\rm bulk}}}\right)-\Delta\rho(\mathbf{r},\mathbf{\omega})\right],\label{eq:1.1}
\end{equation}
where $k_{{\rm B}}T$ is the thermal energy ($\sim0.6$ kcal/mol at 300K), $\textrm{d\ensuremath{\boldsymbol{r}}}\equiv\textrm{dxdydz}$,
$\textrm{d\ensuremath{\omega\equiv}d\ensuremath{\cos\theta}d\ensuremath{\phi}d\ensuremath{\psi}}$
and $\Delta\rho(\mathbf{r},\mathbf{\omega})\equiv\rho(\mathbf{r},\mathbf{\omega})-\rho_{{\rm bulk}}$
the excess density over the bulk homogeneous density.

The external contribution comes from the interaction potential $v_{{\rm ext}}$
between the solute molecule and a solvent molecule. It reads 
\begin{equation}
\mathcal{F}_{{\rm ext}}=\int\mathrm{d}\mathbf{r}\mathrm{d}\mathbf{\omega}\rho\left(\mathbf{r},\mathbf{\omega}\right)v_{{\rm ext}}\left(\mathbf{r},\mathbf{\omega}\right).
\end{equation}
The interaction potential is typically made of a van der Waals term
(Lennard-Jones) and electrostatic interactions. Those are the same
non-bonded force field parameters as in a molecular dynamics simulation.
In what follows, the MDFT computes the SFE for a frozen solute conformer
and therefore does not use intramolecular interactions. 

The final, excess term describes the excess solvent-solvent contribution.
It may be written as a density expansion around the homogeneous bulk
density $\rho_{\textrm{bulk}}$:
    \begin{equation}
        \begin{aligned}
        \mathcal{F}_{{\rm exc}}=&-\frac{k_{\text{B}}T}{2}\int d\mathbf{r}_{1}\mathrm{d}\mathbf{\omega}_{1}\int d\mathbf{r}_{2}\mathrm{d}\mathbf{\omega}_{2}\Delta\rho\left(\mathbf{r}_{1},\mathbf{\omega}_{1}\right)\\
        &\hspace{1.5cm}\times c^{(2)}\left(r_{12},\mathbf{\omega}_{1},\mathbf{\omega}_{2}\right)\Delta\rho\left(\mathbf{r}_{2},\mathbf{\omega}_{2}\right)+\mathcal{F}_{{\rm b}}\\
        =&-\frac{k_{\text{B}}T}{2}\int d\mathbf{r}_{1}\mathrm{d}\mathbf{\omega}_{1}\Delta\rho\left(\mathbf{r}_{1},\mathbf{\omega}_{1}\right)\gamma\left(\mathbf{r}_{1},\mathbf{\omega}_{1}\right)+\mathcal{F}_{\textrm{b}}\\
        =&\hspace{0.1cm}\mathcal{F}_{{\rm HNC}}+\mathcal{F}_{{\rm b}},\\
        \end{aligned}
        \label{eq:excess}
    \end{equation}
where $c^{(2)}\left(r_{12},\mathbf{\omega}_{1},\mathbf{\omega}_{2}\right)$
is the solvent-solvent molecular direct correlation function of the
homogeneous solvent \citep{puibasset_bridge_2012,belloni_2017} with
$r_{12}$ the distance between $\mathbf{r}_{1}$ and
$\mathbf{r}_{2}$. $\mathcal{F}_{\textrm{b}}$ is the bridge functional
and $\gamma\equiv c^{(2)}*\Delta\rho$ the indirect solute-solvent
correlation defined as the spatial and angular convolution of the
excess density with $c^{(2)}$. If one cuts the excess functional expansion at the second order in density, that is, if one cancels the bridge functional\citep{vanleeuwen_1959},
one finds that the MDFT functional produces at its variational minimum
the well-known HNC relation for the solute-solvent distribution function:
\begin{equation}
g(\mathbf{r},\omega)=\frac{\rho_{\text{eq}}(\mathbf{r},\omega)}{\rho_{\textrm{bulk}}}=e^{-\beta v_{\textrm{ext}}(\mathbf{r},\omega)+\gamma(\mathbf{r},\omega)},
\end{equation}
where $\beta\equiv\frac{1}{k_{B}T}$. Therefore, we call the first term of the excess functional the HNC functional. Higher-order correlation functions of the homogeneous reference fluid can be computed but are not numerically tractable as of today.
The rest of the excess term, the so-called bridge functional can be approximated more or less empirically\citep{levesque_scalar_2012,jeanmairet_molecular_2015,gageat_2018}.
However in this paper we benchmark MDFT for small drug-like molecules
at its lowest level of accuracy: the MDFT-HNC. There is no bridge
functional in what follows.

The algorithms to minimize eq. \ref{eq:excess}, are described in Ding et al.\citep{ding_2017}. 
They predict HFEs in few seconds to minutes depending on the supercell size and spacial and angular resolutions. 
We emphesis the importance of including a pressure correction to the 'raw' MDFT-HNC results\citep{sergiievskyi_fast_2014,sergiievskyi_solvation_2015,luukkonen_2020}. 
The following MDFT results include the van der Waals pressure correction described in Robert et al.\citep{robert_2020} and are referred as MDFT$^\mathrm{HNC+vdW}$ results.


\subsection{Computational details}

MDFT calculations are done with a solute embedded in a cubic supercell of length 21 \AA, with periodic boundary conditions, a spatial resolution of 0.33 \AA\quad (= 64x64x64 grid nodes) and an angular resolution of 84 orientations per spacial grid node. 
The MDFT calculations are done on a single frozen configuration of the solute corresponding to the initial configuration given in the FreeSolv database and the solute and solvent molecules are described with the same force field parameters used for the MD+FEP calculations of the FreeSolv database: GAFF force field (v1.7)\citep{wang_gaff_2004} with AM1-BCC partial charges\citep{jakalian_2000,jakalian_2002} for the solutes and the TIP3P\citep{TIP3P} model for the water. 
The average computation time is 1 min 53 sec on a single CPU-thread. The MDFT minimization process did not converge for 23 solutes (4\% database, see SI for more information). 
All results we present below are for the 619 molecules that converged. 


\section{Benchmarking MDFT for small drug-like molecules}

\begin{figure}[h!]
\begin{centering}
\includegraphics[width=9cm]{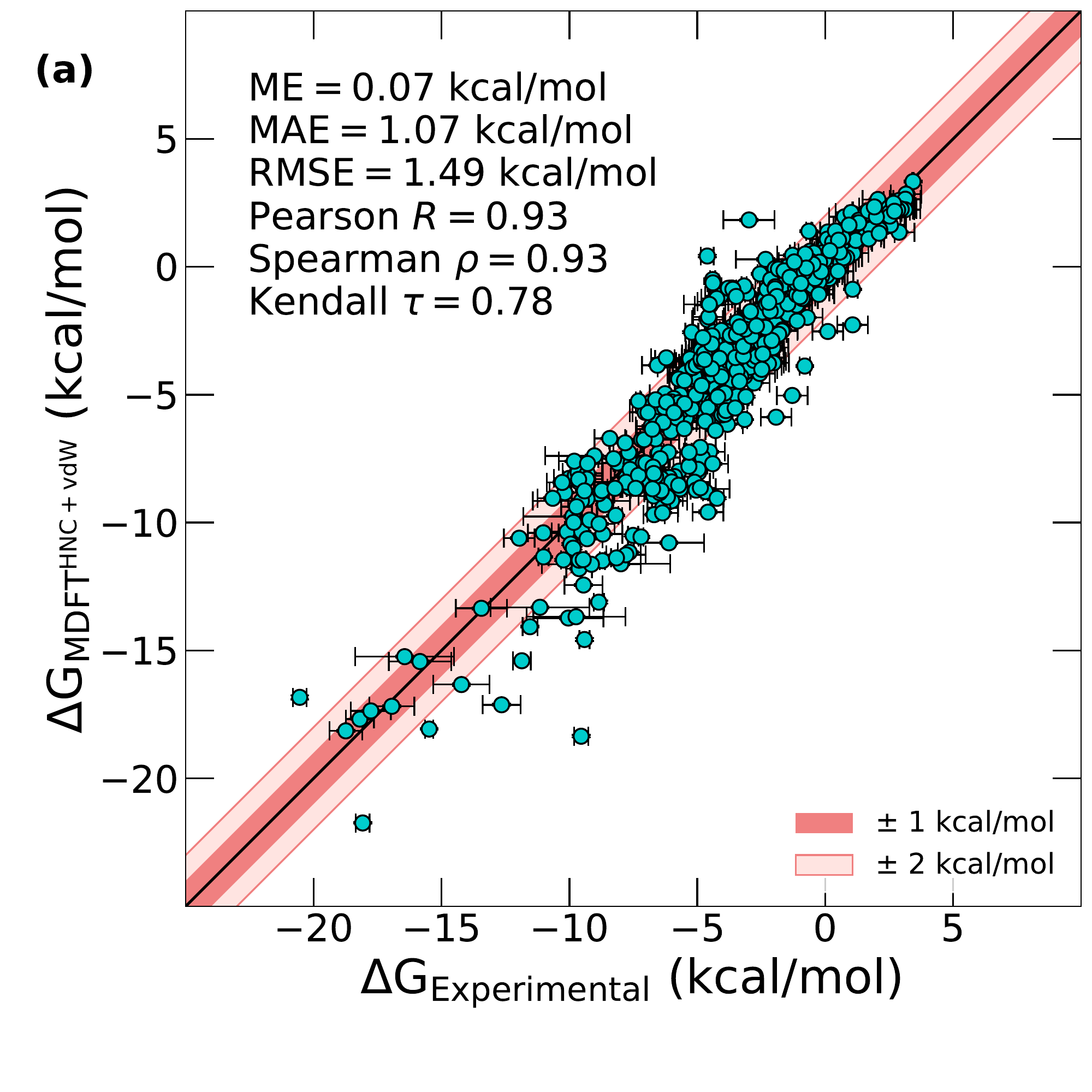} \\
\includegraphics[width=9cm]{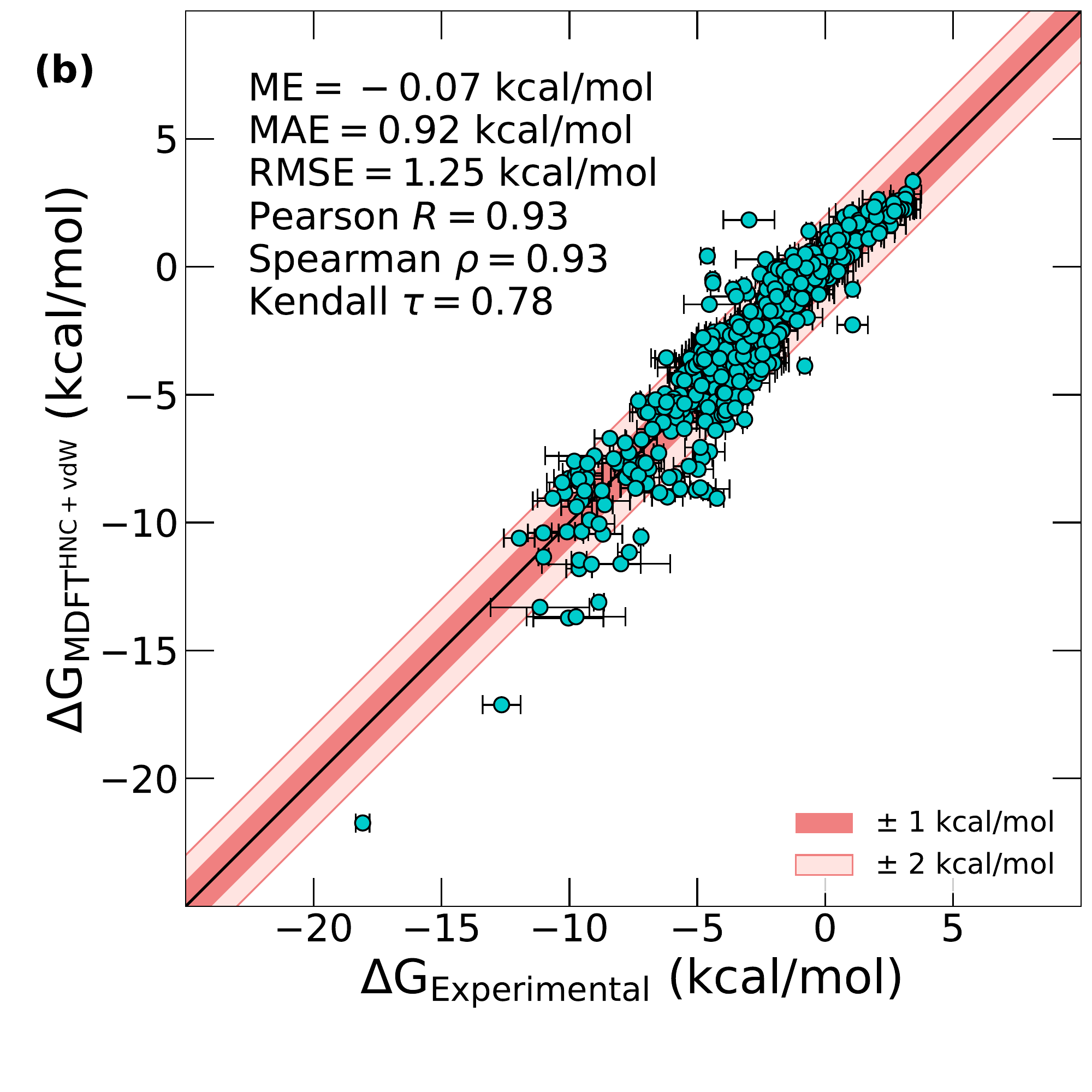}
\par\end{centering}
\centering{}\caption{Correlations between hydration free energies predicted by MDFT-HNC with the van der Waals pressure correction and those measured experimentally for (\textbf{a}) the whole FreeSolv
database (619 molecules) and (\textbf{b}) a subset of rigid molecules of the FreeSolv database (520 molecules). The subset is defined in the text. \label{fig:correlations-full}}
\end{figure}

In figure \ref{fig:correlations-full}a, we show the correlation between
experimental HFEs and those obtained with $\mathrm{MDFT^{HNC+vdW}}$. The mean absolute error (MAE) is 1.07 \textpm{} 0.08 kcal/mol and the Pearson's correlation
coefficient $R$ is 0.93 \textpm{} 0.01. MDFT results also have a
small mean (signed) error (ME) of 0.07 \textpm{} 0.12 kcal/mol which
indicates that MDFT does not have a systematic bias : it is lower
in amplitude than the statistical error bars. All the statistical
measures characterizing this correlation are summarized in table \ref{tab:Summary-of-the}.
The error bars on the measures correspond to the 95\% confidence interval\citep{errorbars}. Note that as we here are comparing $\mathrm{MDFT^{HNC+vdW}}$, an approached theory, to experimental data, the deviations could be the results of incorrect approximations in $\mathrm{MDFT^{HNC+vdW}}$ or due to bad force field parameterization.

\begin{figure*}[h!]
\includegraphics[width=19cm]{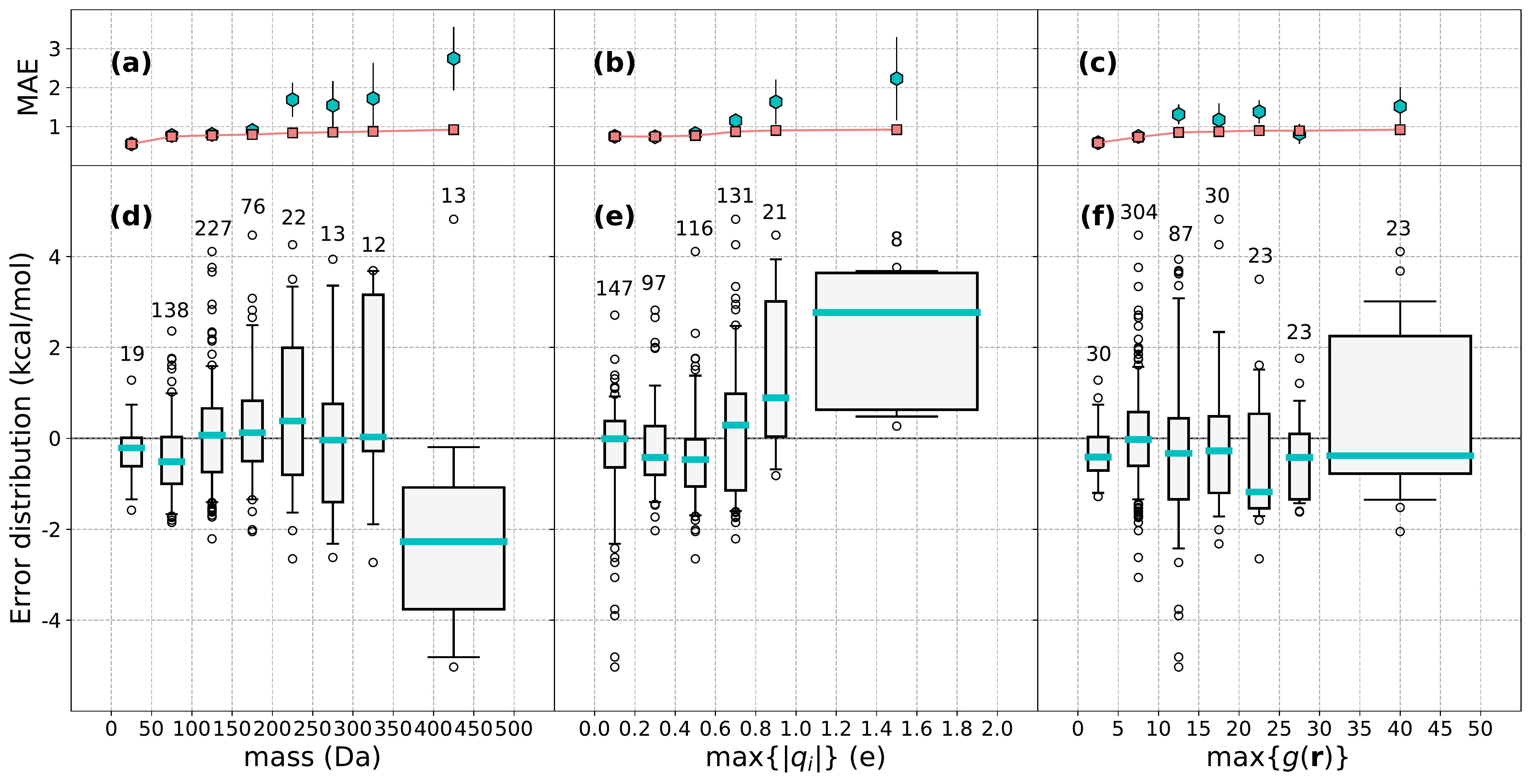}\caption{Distribution of mean absolute error between predicted and experimental hydration free energies as a function of three features (\textbf{a}) solute's molar mass (with bin size of 50 Da), (\textbf{b}) solute's largest local charge (0.2 e) and (\textbf{c}) the maximum of the 3D $g$ around the solute (5) with turquoise hexagons (error bars correspond to the 95\% confidence interval) presenting the MAE
of each bin and pink squares the cumulative MAE. 
The corresponding signed error distributions are presented in \textbf{d},\textbf{e} and \textbf{f} with turquoise lines corresponding to median error in each bin, the boxes and the whiskers to 25-75\% and 5-95\% intervals respectively and black circles to fliers outside the 5-95\% interval.
The numbers above each bin is the population of the bin for the FreeSolv database and for each distribution the last three (350-500 Da), five (1-2 e) or four (30-50) bins are gathered into one in order to be statistically significant.
\label{fig:Error-distributions-along}}
\end{figure*}

Table \ref{tab:Summary-of-the} contains also the statistical measures
characterizing the correlations between experimental values and those
obtained with MD+FEP given by Duarte Ramos Matos et al.\citep{matos_2017_freesolv}
and those obtained with 3D-RISM-KH by Roy and Kovalenko\citep{roy_2019}.
Overall the three methods perform at the same accuracy level with
similar errors and correlation coefficients. However, MDFT's computation
time is on average less than 2 cpu.min compared to hundreds cpu.h
or tens gpu.h with MD+FEP and few tens cpu.min with 3D-RISM\citep{palmer_towards_2010,rism-time}. Hence for the same accuracy, MDFT has a speedup of 1-2 and 3-4 orders
of magnitude when compared to 3D-RISM and MD+FEP respectively. Compared to 3D-RISM, MDFT does not have the consequences from approximating MOZ\citep{hansen_theory_2013,sullivan_1981,Chandler_1978,morriss_1981}.


\subsection{Effect of flexibility}

Solute flexibility can have an important effect on HFEs\citep{mobley_2008,klimovich2010_sampl2}, however as mentioned before, the current MDFT calculation is done on a single rigid conformation of the solute, hence the solute flexibility is not taken into account in our MDFT calculation.
Therefore, we studied a subset of 520 quasi-rigid solutes to estimate
the importance of the lack of solute flexibility in MDFT. 
We define a solute as rigid if the HFEs obtained with flexible MD+FEP\citep{matos_2017_freesolv} and rigid solute MC+FEP simulations\citep{belloni_2019,robert_2020} differ by less than 0.6 kcal/mol which is the average experimental error of the database. 

We show the correlation between experimental HFEs and MDFT results for this subset of rigid molecules in figure \ref{fig:correlations-full}b and the statistical measures characterizing the correlations are summarized in table \ref{tab:Summary-of-the}. 
We observe only a slight improvement of the MAE ($-14\%$) and the RMSE ($-16\%$). However, most of MDFT's largest outliers can be attributed to the single conformer approximation of the method as the maximum error decreases from 8.81 to 4.82 kcal/mol ($-45\%$) and the number of solutes with absolute errors larger than 3 kcal/mol decrease from 39 to 17 ($-56\%$) when limiting ourselves to rigid solutes.

The following error analyses are done on the subset of 520 rigid solutes.


\subsection{Effect of solute's mass, charges and solvation structure}

In order to give an optimal set of requirements and confidence intervals to MDFT-HNC predictions, we now focus on finding sources of errors or correlations between errors.
Figure \ref{fig:Error-distributions-along} shows the error distribution in function of the solute's (i) molar mass, (ii) largest partial charge $\mathrm{max}\{|q_{i}|\}$ and (iii) highest value of the 3D solvation structure $\mathrm{max}\{g(\boldsymbol{r})\}$. 
As shown in figure \ref{fig:Error-distributions-along}a, the heaviest molecules have the largest deviations to experiment: the MAE increases with the solute's mass. 
For solutes with a molar mass larger than 200 Da the MAE is 1.75 kcal/mol, i.e. almost the double than for the whole database.
However these molecules present only 12\% of the rigid subset so their effect on the total MAE is not significant as seen on the cumulative MAE. 
Similar trends are present also for the MD+FEP and RISM results with a MAE of 1.78 and 2.21 kcal/mol respectively for these molecules larger than 200 Da (see Figure S1).

Similarly to the molar mass, the deviation to experiment increases with the magnitude of the largest partial charge of the drug-like molecule, positive or negative, (see fig. \ref{fig:Error-distributions-along}b) with a MAE of 1.84 kcal/mol for solutes with $\mathrm{max}\{|q_{i}|\}>0.8e$ (6\% of the rigid subset). 
The effect is less pronounced for MD+FEP and RISM with MAEs at 1.55 and 1.50 kcal/mol respectively for these molecules (see Figure S1). 
This is expected for MDFT at the HNC approximation: the second order density expansion of the functional around $\rho=\rho_{\mathrm{bulk}}$, or $g=1$, misses higher order repulsion terms.
This leads to problems for cases with densities getting away from $\rho_{\mathrm{bulk}}$: either high densities typically found next to high (partial) charges or large solutes with large volumes where $g=0.$ 

\begin{figure}[h!]
\includegraphics[width=9cm]{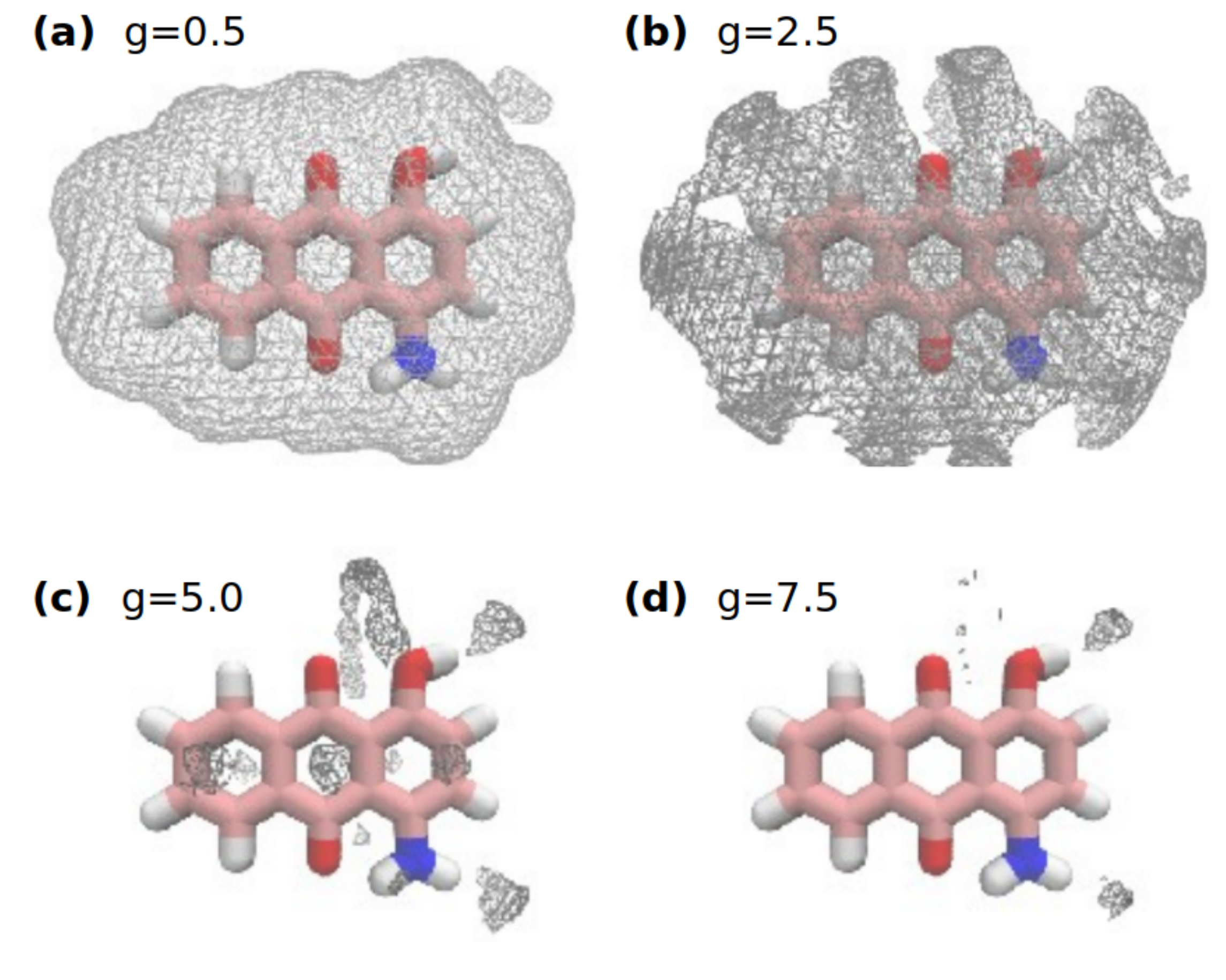}
\caption{Water density map isosurfaces around 1-amino-4-hydroxy-9,10-anthracenedione
at \textbf{a.} $g=0.5$, \textbf{b.} $2.5$, \textbf{c. $5.0$ }and \textbf{d. }$7.5$. \label{fig:isosurfaces}}
\end{figure}

Besides the solute's molar mass and partial charges, solute features known a priori, we can also look at the output of a MDFT calculation, that is, the solvation profile, to predict, on this dataset at least, the quality of the MDFT's HFE predictions. 
In figure \ref{fig:isosurfaces}, we illustrate the 3D solvent density around 1-amino-4-hydroxy-9,10-anthracenedione (FreeSolv ID: 4371692) with four water-oxygene density isosurfaces ($g=0.5,2.5,5.0$ and $7.5$). Water density maps, that are time consuming to compute using MD\citep{coles_2019} are a direct output of MDFT, again in obtained in 2 min on average.
Low densities (fig. \ref{fig:isosurfaces}a) are observed on the limits of the solute's cavity but also after the first solvation peak (fig. \ref{fig:isosurfaces}d) of the hydroxyl group. The largest oxygen densities (fig. \ref{fig:isosurfaces}d) are observed next to the hydroxyl-hydrogen and the less crowded amine-hydrogen that are potential hydrogen-bond donors.

In, figure \ref{fig:Error-distributions-along}c. we see that the MDFT's deviation to experiment increases with the maximum height of the solvation peaks with a MAE of 1.24 kcal/mol for solutes with $\mathrm{max}\{g(\boldsymbol{r})\}>20$ (1\% of the rigid subset). 
This result is expected as high density peaks are difficult cases for the HNC approximation as discussed in the previous paragraph. 
However, the link between the amplitude of the deviation and the solvation structure is less pronounced as for the solute's mass and partial charges. 

\begin{figure*}[h!]
\includegraphics[width=19cm]{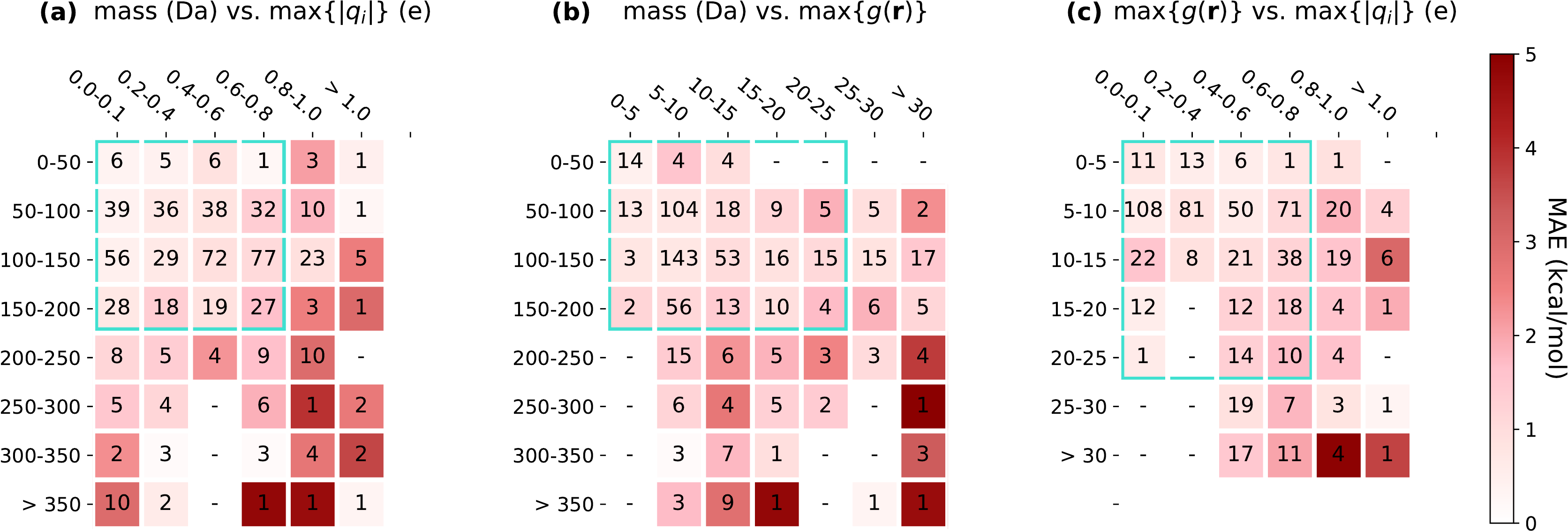}
\caption{Distribution of mean absolute error between predicted and experimental hydration free energies as a function of two features: \textbf{a.} solute's
mass and largest partial charge, \textbf{b.} solute's mass and highest
solvation peak and \textbf{c.} solute's highest solvation peak and
largest partial charge. The numbers in each bin is the population
of the bin. \label{fig:Cross-err}}
\end{figure*}

In figure \ref{fig:Cross-err}, we show two-dimensional cross distributions of MAE for the three features studied above. 
Often solutes with high mass and high charges/solvation peaks have the largest deviations but deviations can be large for molecules with only one feature with a 'high' value (eg. MAE=3.05 kcal/mol for solutes with $\mathrm{max}\{g(\boldsymbol{r})\}=10-15$ and $\mathrm{max}\{|q_{i}|\}>1.0e$ in \ref{fig:Cross-err}c). 
The smallest deviations from experiment are found for the solutes with a mass of less than 200 Da, largest partial charge of less than 0.8 and highest solvation peak at less than 25, delimited by the turquoise rectangles in figure \ref{fig:Cross-err} (73\% of the database), with a MAE at $0.73\pm0.22$ kcal/mol. A table of the three-dimensional
cross distributions of MAE is given in SI.

\subsection{Effect of functional groups}

\begin{figure*}[h!]
\includegraphics[width=19cm]{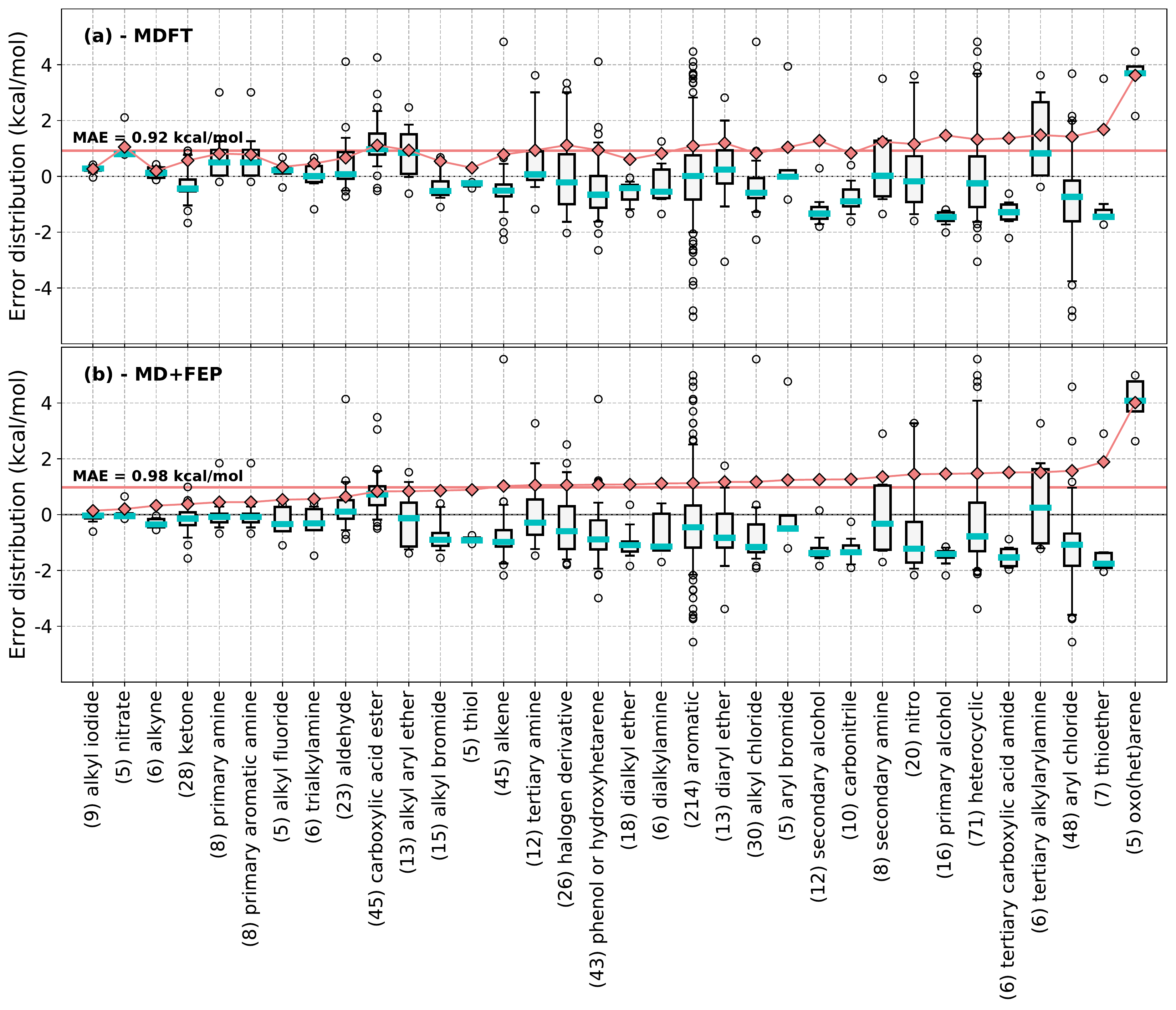}\caption{Error distributions of (a) MDFT$^\mathrm{HNC+vdW}$ and (b) MD+FEP for the chemical groups with more than 5 solutes present in the database. The number of molecules in each group is written within parenthesis. Turquoise lines correspond to median error in each bin, the boxes and the whiskers to 25-75\% and 5-95\% intervals respectively and black circles to fliers outside the 5-95\% interval. Pink diamonds correspond to the MAE of each functional group and the vertical pink line to total MAE of the rigid subset. \label{fig:Mean-MAE-by-group}}
\end{figure*}

Here we asses the performance of MDFT as a function of the chemical groups present in a solute.
In figure \ref{fig:Mean-MAE-by-group} we show the error distribution of MDFT$^\mathrm{HNC+vdW}$ and reference MD+FEP as a function of each chemical function present in at least five molecules of the database.

We observe a high correlation between the MAEs of MDFT$^\mathrm{HNC+vdW}$ and MD+FEP ( $R=0.90$ and $\rho=0.81$). In general functional groups with small/large errors with MD+FEP also have small/large errors with MDFT$^\mathrm{HNC+vdW}$. This indicates that the major part of MDFT's error comes from the force field parametrization and not the approximated theory itself. This is not unexpected sinci it was shown that MDFT with appropriate partial molar volume corrections reproduce similar SFE's with an accuracy of $k_BT$\citep{luukkonen_2020} or below\citep{robert_2020}. For example it has been noted that the GAFF parametrization of the hydroxyl groups leads to systematic errors in HFEs computed with MD+FEP\citep{fennell_2014}. Here we observe above average MAEs for primary and secondary alcohols with systematic underestimation of the HFEs (ME < 0 with narrow distribution of errors) for both MD+FEP and MDFT$^\mathrm{HNC+vdW}$.

Nonetheless, there are differences between MDFT's and MD+FEP's MAEs:  MDFT$^\mathrm{HNC+vdW}$ significantly over-performs some groups, like thiols, or under-performs for other groups, like nitrates, when to compared to MD+FEP. Hence the totality of MDFT's error cannot be attributed to the force field parametrization.

\begin{figure}[h]
\includegraphics[width=9cm]{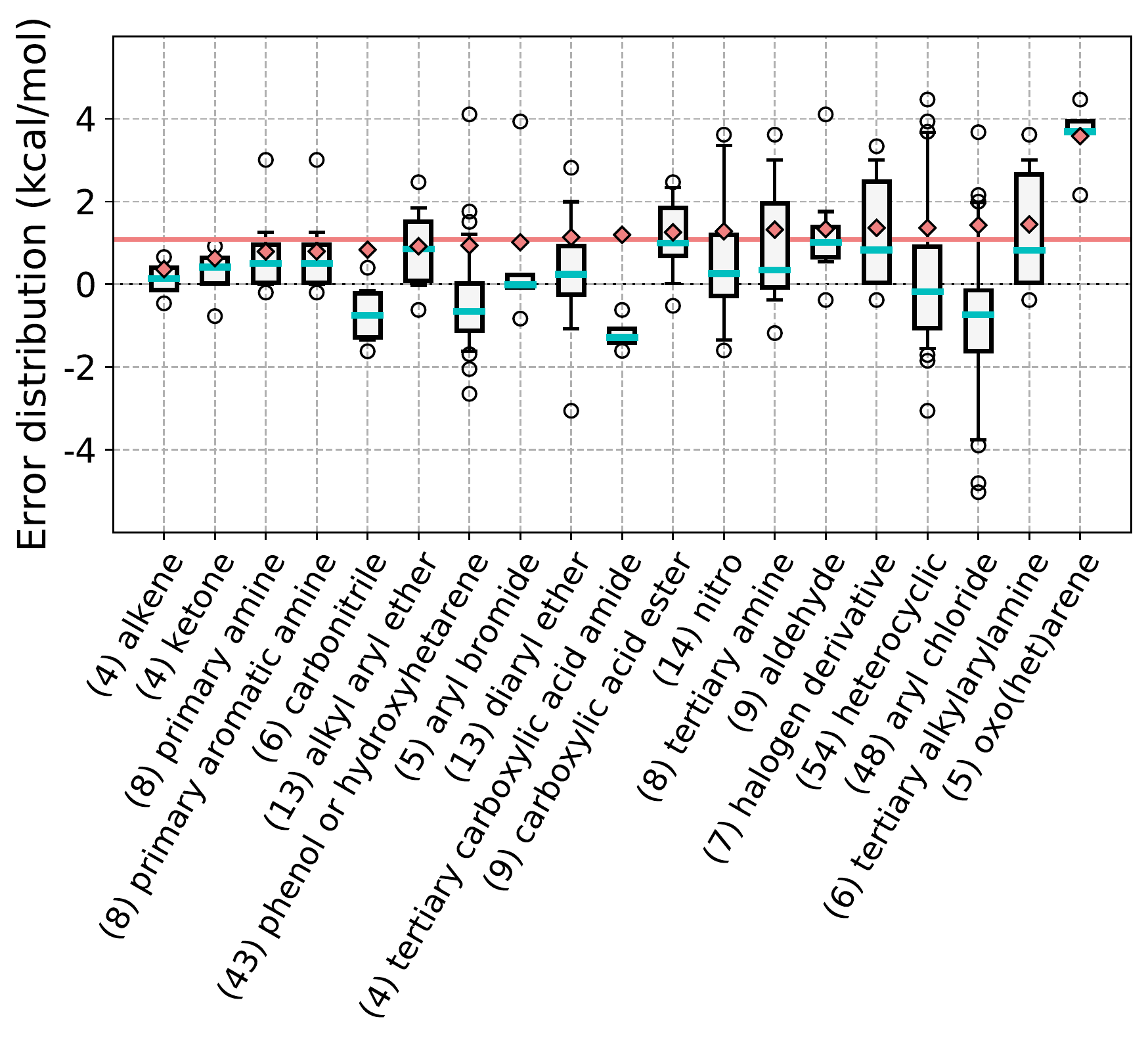}
\caption{Error distributions of solutes with an aromatic ring coupled to another chemical function for couples present in more than five solutes. The number of molecules in each couple is written within parenthesis. Turquoise lines correspond to median error in each bin, the boxes and the whiskers to 25-75\% and 5-95\% intervals respectively and black circles to fliers outside the 5-95\% interval. Pink diamonds correspond to the MAE of each functional group and the vertical pink line at 1.18 kcal/mol corresponds to the MAE of all the molecules containing an aromatic ring.
\label{fig:Aromatics}}
\end{figure}

Additionally, we did a similar cross-analysis between chemical functions, as for the mass, partial charge and solvation peak couples. 
A table of all error bars reconstructed from this analysis is given in SI.
To illustrate these error estimates, in figure \ref{fig:Aromatics} we show the distribution of MAE for molecules with an aromatic ring, the most frequent chemical function in the database (present in 214 solutes, i.e. 41\% of the rigid subset), coupled with another chemical group. We see that in most cases the MAE of an aromatic+another group is close to the overall MAE of the aromatics.

Most notably exception is the MAE of aromatic+oxo(het)arene at 3.58 kcal/mol which is much higher than the MAE of all aromatics at 1.08 kcal/mol. 
This is coherent with oxo(het)arenes having the largest errors of all functional groups. More interesting are the couples like aromatic+alkene (MAE=0.37 kcal/mol) or aromatic+aldehyde (MAE=1.34 kcal/mol) for which the MAEs of the couples are lower or higher than the MAE of the individual chemical functions that they are composed of. Note that these couples contain only 2 and 4 solutes each so these behaviours might be artifacts of limited sampling.

 
\section{Conclusion }

Molecular density functional theory in the hyper-netted chain approximation coupled with revised pressure correction predicts experimental hydration free energies with a mean absolute error of 1.07 kcal/mol in 2 minutes on average. 
Experimental values were available to us during this work but no fitting or adjustments were done to the MDFT method to reproduce the experimental values\footnote{The van der Waals pressure correction\citep{robert_2020} was fitted
on simulations.}. 
Moreover, for rigid solutes MDFT's accuracy is  below 1 kcal/mol. 
Overall MDFT is at the same level of accuracy as MD+FEP or RISM. 
MDFT takes on average 2 cpu.min per solute to compute HFE compared to 10+ gpu.h (or 100s cpu.h) by MD+FEP giving a speedup of 3-4 orders of magnitude. 
As MD+FEP is exact in force field approximation, and as MDFT and MD+FEP have the same accuracy and as the error distributions in function of the chemical functions are similar, the major source of error in MDFT predictions is the force field parameterization. 

Looking at the solute's molar mass, partial charges and the MDFT solvation profile we can estimate the quality of the MDFT HFE prediction. 
For solutes with a molar mass of less than 200 Da, the largest partial charges at less than 0.8e and highest solvation peak at less than 25, MDFT's MAE is 0.75 kcal/mol (73\% of the database). 
In this work, we also extracted error bars for \{mass, partial charge, solvation peak\} triplets and for each kind of chemical function from which we are now able to infer confidence intervals.
Additionally, our group is working on coupling MDFT with machine learning approaches to improve MDFT's accuracy. The error distribution analysis gives information on the types of molecules for which the corrective machine learning coupling will be important and more reference data needs to be produced.

\subsection{Acknowledgments}

This work has been supported by the Agence Nationale de la Recherche, Project No. ANR BRIDGE AAP CE29.

\subsection{Supporting Information Available:}

Table S1 : Information on solutes that did not converge with MDFT. \\
Table S2 : MDFT error bars as a function of the solute's chemical functions. \\
Table S3 : MDFT error bars as a function of the solute's mass, charge and solvation peak. \\
Figure S1 : Error distributions as a function of the solute's mass for MD+FEP and 3D-RISM. \\

\bibliography{biblio}

\begin{mcitethebibliography}{67}
\bibitem[err()]{errorbars}
For each statistical measure {X} characterizing a dataset of {N} points (eg.
  {N}=619 for the full set), the measure {X}' was computed 10 000 times on {N}
  values chosen at random each time from the dataset. The error bars of {X}
  correspond to two standard deviations of the {X}' distribution.\relax
\mciteBstWouldAddEndPunctfalse
\mciteSetBstMidEndSepPunct{\mcitedefaultmidpunct}
{}{\mcitedefaultseppunct}\relax
\EndOfBibitem
\end{mcitethebibliography}

\end{document}